\documentclass[aps,prb,amsmath,amssymb,reprint,superscriptaddress,floatfix,showpacs]{revtex4-1}

\usepackage{graphicx}% Include figure files
\usepackage{dcolumn}% Align table columns on decimal point
\usepackage{bm}% bold math
\usepackage{hyperref}% add hypertext capabilities
\usepackage[dvipsnames]{xcolor}

\usepackage[version=3]{mhchem} % Formula subscripts using \ce{}
\usepackage[T1]{fontenc}       % Use modern font encodings

\begin{document}

\title{Stability of Long-lived Antiskyrmions in Mn--Pt--Sn Material}

\author{M. N. Potkina}
\affiliation{Science Institute and Faculty of Physical Sciences, University of Iceland, 107 Reykjav\'{\i}k, Iceland}
\affiliation{Department of Physics, St. Petersburg State University, 198504 St. Petersburg, Russia}
\affiliation{Faculty of Physics and Engineering, ITMO University, 197101 St. Petersburg, Russia}
\author{I. S. Lobanov}
\affiliation{Department of Physics, St. Petersburg State University, 198504 St. Petersburg, Russia}
\affiliation{Faculty of Physics and Engineering, ITMO University, 197101 St. Petersburg, Russia}
\author{O. A. Tretiakov}
\affiliation{School of Physics, The University of New South Wales, Sydney 2052, Australia}
\author{H. J\'onsson}
\affiliation{Science Institute and Faculty of Physical Sciences, University of Iceland, 107 Reykjav\'{\i}k, Iceland}
\affiliation{Department of Applied Physics, Aalto University, FIN-00076 Espoo, Finland}
\author{V. M. Uzdin}
\affiliation{Department of Physics, St. Petersburg State University, 198504 St. Petersburg, Russia}
\affiliation{Faculty of Physics and Engineering, ITMO University, 197101 St. Petersburg, Russia}

%\eads{\mailto{hj@hi.is, v_uzdin@mail.ru}}

%%%%%%%%%%%%%%%%%%%%%%%%%%%%%%%%%%%%%%%%%%%%%%%%%%%%%%%%%%%%%%%%%%%%%
%

\begin{abstract}
The lifetime of antiskyrmions at room temperature in a Mn--Pt--Sn tetragonal Heusler material has been calculated using an atomic scale representation including nearly a million spins. The evaluation of the pre-exponential factor in the Arrhenius rate expression for this large system is made possible by an implementation of harmonic transition state theory that avoids evaluation of the eigenvalues of the Hessian matrix. The parameter values in the extended Heisenberg Hamiltonian, including anisotropic Dzyaloshinskii-Moriya interaction, are chosen to reproduce experimental observations  [A.\,K. Nayak \textit{et al.}, Nature {\bf 548}, 561 (2017)], in particular the 150 nm diameter. The calculated results are consistent with the long lifetime observed in the laboratory and this exceptional stability of the antiskyrmions is found to result from large activation energy for collapse due to strong exchange coupling while the pre-exponential factor in the Arrhenius expression for the lifetime is found to have a typical magnitude of 10$^{-12}$\,s, despite the large number of spins.  The long lifetime is, therefore, found to result from energetic effects rather than entropic effects in this system.
\end{abstract}

\maketitle

%%%%%%%%%%%%%%%%%%%%%%%%%%%%%%%%%%%%%%%%%%%%%%%%%%%%%%%%%%%%%%%%%%%%

\section{Introduction}

%Localized, non-collinear magnetic states, in particular 
Skyrmions and antiskyrmions are localized magnetic states that have been proposed as elements in
future spintronics devices.\cite{Kiselev_2011,Fert_2013,Fert_2017}
Along with interesting transport properties, such states can exhibit particle-like behavior and carry a topological charge that enhances their stability with respect to the uniform ferromagnetic or antiferromagnetic states.
A key issue is the lifetime of (anti)skyrmions and its dependence on temperature and applied magnetic field.
The challenge is to find or design materials where such magnetic states are sufficiently stable at ambient temperature 
and still small enough to be used in high density spintronic devices. 
Figure 1 shows the spin configuration of an antiskyrmion as well as that of a skyrmion. 

So far, stability at room temperature has mainly been obtained for large skyrmions with a diameter of 50 nm or more.\cite{Everschor-Sitte_2018,Soumyanarayanan_2017} 
%The challenge is to find materials with smaller (anti)skyrmions that are still stable enough at room temperature.
%For this purpose 
It is important to understand what determines the lifetime in order to 
guide the search for materials where smaller (anti)skyrmions are sufficiently stable at room temperature.
%  change in resubmission
%are primarily stabilized by entropy rather than the energy barrier for collapse.
From recent experimental studies of skyrmions in Fe$_{1-x}$Co$_x$Si, 
a large, destabilizing entropic contribution which reduces the lifetime of skyrmions has been reported.
\cite{Wild_2017}
On the other hand, theoretical studies have found that isolated skyrmions can be stabilized by entropic contributions.
\cite{Varentsova_2018,Desplat_2018,Malottki_2019,Desplat_2020,Hoffmann_2020}
The question we address here is whether this is also the case for the recently observed stable antiskyrmions in 
acentric tetragonal Heusler compounds.\cite{Nayak_2017}
The diameter of these antiskyrmions is large, 150 nm, 
%and the question is whether materials parameters could be modified in 
but it may be possible to modify materials parameters in
some way to obtain smaller antiskyrmions that are still stable at room temperature. 
%Another question we raise is how antiskyrmions compare with skyrmions in terms of stability. 
Antiskyrmions offer some advantage over skyrmions in that they can under some conditions move in the direction of an applied spin polarized current,
while skyrmions necessarily move at an angle.\cite{Huang_2017}
%The two are related by a transformation. 

Skyrmions have been studied for several systems and
recent reviews have summarized results obtained.\cite{Fert_2017,Everschor-Sitte_2018}
The annihilation of a skyrmion can occur through various mechanisms, in particular 
collapse in the interior of the sample\cite{Bessarab_2015,Lobanov_2016,Stosic_2017,Uzdin_2018} 
and escape through the boundary of the magnetic domain.\cite{Stosic_2017,Uzdin_2018,Bessarab_2018} 
Two skyrmions can also merge to form a single skyrmion (and the reverse can also occur, i.e. a division of a skyrmion into two).\cite{Muller_2018}
The calculated lifetime estimates have taken into account the influence of a magnetic field\cite{Bessarab_2018}, point defects,\cite{Uzdin_2018}
and the width of the track where the skyrmion resides.\cite{Uzdin_2018} 
Calculations have also been carried out for skyrmions in antiferromagnets.\cite{Bessarab_2019,Potkina_2020} 
Recent calculations have addressed how the various materials parameters, such as
the Dzyaloshinskii-Moriya interaction (DMI) and anisotropy constants affect the 
activation energy for the collapse of a skyrmion.\cite{Varentsova_2018}
The results show that the activation energy is not simply a function of the size of the skyrmion although the two tend to be correlated.
%Theoretical calculations of the lifetime of large and stable skyrmions, where the number of spins is of the order of $10^{6}$, are, however, challenging and have not been reported.

Fewer studies have been carried out on antiskyrmions.
%Recently, antiskyrmions in acentric tetragonal Heusler compounds with diameter of 0.15 microns have been found to be stable at room 
%temperature\cite{Nayak_2017}. 
They are stabilized by 
anisotropic Dzyaloshinskii-Moriya interaction (DMI) while isotropic DMI stabilizes skyrmions.\cite{Meshcheriakova_2014, Nayak_2017}
Some aspects of antiskyrmions in systems with anisotropic DMI 
have been studied theoretically.\cite{Gungordu_2016, Hoffmann_2017, Huang_2017, Kovalev_2018}
Antiskyrmions 
%of similar size 
can also be stabilized by the magnetostatic interactions, for example in 
ion-irradiated Co/Pt multilayer films\cite{Phatak_2016}.  
Antiskyrmions have, furthermore, been discussed in relation 
to skyrmion--antiskyrmion pair production,\cite{Koshibae_2014, Koshibae_2016, Stier_2017}
in particular in frustrated ferromagnetic films.\cite{Leonov_2015, Malottki_2017, Zhang_2017}
Monte Carlo simulations using parameters estimated from electronic density functional theory calculations
have been used to simulate both skyrmions and antiskyrmions in the Pd/Fe/Ir(111) system
and the predicted spin-polarized scanning tunneling microscopy images found to be similar.\cite{Dupe_2016}
Antiskyrmions as well as skyrmions in ferromagnetic films have been simulated using the micromagnetic approximation
and the effect of the dipole-dipole interaction shown to provide larger stabilization for antiskyrmions.\cite{Camosi_2018}
The energy barrier for collapse cannot, however, be evaluated within the micromagnetic approximation since the collapse mechanism involves a
singularity that requires a discrete lattice representation.

It is of great interest for potential spintronic applications to carry out more studies of antiskyrmions. 
Only a few calculations of antiskyrmions have been reported so far, 
and they have mainly focused on stabilization by frustrated exchange rather than the anisotropic DMI.\cite{Malottki_2017,Ritzmann_2018,Desplat_2019}
%Calculations using a micromagnetic description
%have shown that the dipole-dipole interaction stabilizes the antiskyrmion with respect to the skyrmion\cite{Camosi_2018}
%but the energy barrier for collapse cannot be evaluated within the micromagnetic approximation. 
Simulation studies of the factors that affect thermal stability have, furthermore, been limited so far to rather small skyrmions that are unstable at room 
temperature. 
%  change in resubmission
%The 
An important challenge is to extend the simulation methodology in order to make it applicable to large enough lattices  
to accurately represent the large (anti)skyrmions that have been found experimentally to be stable at room temperature. 

In this article, the stability of antiskyrmions is evaluated using an atomic scale representation and harmonic transition state theory.
The calculations reproduce large antiskyrmions observed in the Mn--Pt--Sn inverse Heusler compound.\cite{Nayak_2017}
The long lifetime at room temperature is found to be due to high energy barrier for collapse resulting mainly from strong exchange interaction.
%in the presence of the anisotropic DMI.
The pre-exponential factor which 
%  change in resubmission
%represents 
includes
the entropic effects is, however, found to be of a typical magnitude, 10$^{-12}$\,s, 
despite the large number of spins involved.

% =============================================================================

% -----------------------------   Figure 1 ----------------------------------
\begin{figure}[t]
\centering
\includegraphics[width=8 true cm]{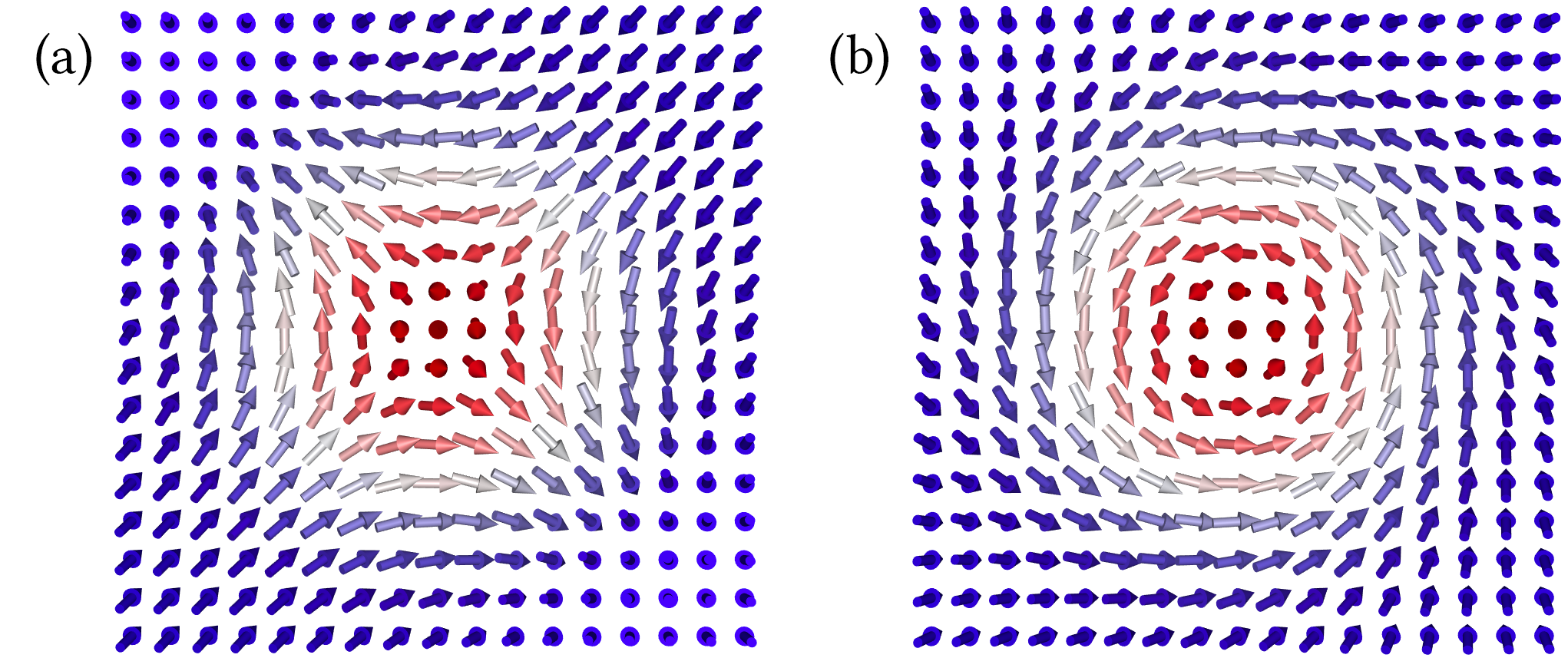}
\caption{\label{fig:1} 
(a) Antiskyrmion.
(b) Bloch skyrmion.
The color (red vs. blue) indicates the direction of the out-of-plane component of the
magnetic vector and the intensity of the color indicates the magnitude.
}
\end{figure}
% -------------------------------------------------------------------------------

\section{Simulated system}

The system is 
described by a Heisenberg-type Hamiltonian 
\begin{equation}
\label{eq:Hamiltonian}
\mathcal{H} = - \sum_{\langle i, j\rangle}\! \left[ J \mathbf{S}_i \cdot \mathbf{S}_j + \mathbf{D}_{ij} \cdot (\mathbf{S}_i \times \mathbf{S}_j ) \right]
-   \sum_{i} [\mu  \textbf{B} \cdot \mathbf{S}_i + K S_{i,z}^2]. 
\end{equation}			
Here $J$ is the exchange constant for the nearest neighbor magnetic moments ($J >0$), 
$\mathbf{D}_{ij}$ is the Dzyaloshinskii-Moriya vector lying in the plane of the sample ($x$-$y$ plane), 
$K$ is the uniaxial anisotropy constant, $\textbf{B}$ is the magnetic field, and $\mathbf{S}_i$ is the vector of unit length in the direction of the magnetic moment at site $i$ of a square lattice. The summation $\langle i, j \rangle$ 
runs over pairs of nearest neighbor sites.
We note that the dipole-dipole interaction is not included in the Hamiltonian in the present calculations.
%The spins at the boundary of the simulated system are free.  
Free boundary conditions are used resulting in
an energy barrier for the escape of the antiskyrmion through the boundary and eliminating
%free translation 
translational invariance
that would lead to zero modes. Such modes require special treatment in the lifetime calculation.\cite{Ivanov_2017}

Depending on the type of DMI, 
this Hamiltonian can give rise to skyrmions (Bloch or N\'eel) or antiskyrmions.
If the DMI vector points along the bond connecting sites $i$ and $j$, a Bloch type skyrmion 
% new in resubmitted manuscript
(shown in fig.1(b)) 
can form. 
The vector can be written as $\mathbf{D}_{ij}  = ({ \hat {r}}_{ij}\cdot {\hat {x}}) \mathbf{D}_1 + ({\hat {r}}_{ij}\cdot { \hat {y}}) \mathbf{D}_2$ 
where ${\hat r}_{ij}$ is a unit vector pointing from site $i$ to site $j$. 
An anisotropic DMI with $\mathbf{D}_1 = (D,0,0)$ and $\mathbf{D}_2 = (0,-D,0)$ in Eq.~(\ref{eq:Hamiltonian})
supports a symmetric antiskyrmion 
% new in resubmitted manuscript
(shown in fig.1(a)).
\cite{Bogdanov_1989,Gungordu_2016}

The parameters in the extended Heisenberg Hamiltonian are chosen here to mimic the Mn--Pt--Sn inverse Heusler compound
where antiskyrmions have been observed over long time scale at room temperature.\cite{Nayak_2017}
The diameter of an antiskyrmion in this material has been measured to be approximately 150 nm, corresponding to 230 lattice constants, 
and accurate atomic scale modeling therefore requires a square lattice containing at least $900\times900$ lattice points, i.e. nearly a million spins.
In order to evaluate the activation energy and estimate the lifetime of the antiskyrmion, it is necessary to use a discrete atomic lattice, 
rather than the continuum approximation invoked in micromagnetic simulations.

The parameters in the Hamiltonian used here are chosen to be consistent with the previously determined micromagnetic model parameters for this system.\cite{Nayak_2017}
There, the
%micromagnetic simulations in Ref. \cite{Nayak_2017} were found to correspond to the experimentally observed 
antiskyrmions at $T=300$ K
were modeled using the following parameter values: 
exchange stiffness $A=1.2\times 10^{-10}$ J/m,  
Dzyaloshinskii-Moriya parameter $d= 6\times 10^{-3}$ J/m$^2$, 
saturation magnetization $M_s = 445$ kA/m,  
external field $B=0.29$ T, 
zero anisotropy, and an in-plane lattice constant of $a=$ 0.63 nm and out-of-plane lattice constant of $c=$ 1.22 nm.
These micromagnetic parameter values are converted to parameters for the atomic scale lattice Hamiltonian 
as $J=2 c A =1830$ meV, 
$D=a c d =29$ meV, $K=0$, and $\mu=a^2 c M_s =1.37$ meV/T. 
We note, however, that the micromagnetic simulations can only be used to determine these quantities relative to the exchange parameter, $J$,
so our calculations are carried out in terms of scaled parameters
$D/J=0.016$\ and $\mu/J=7.5\cdot$10$^{-4}$ T$^{-1}$.

% ---------------------------

\section{Calculations of the lifetime} 

The lifetime of a magnetic state can generally be described by an Arrhenius rate law, $\tau =\tau_0 \exp(E_a/k_B T)$, where 
$E_a$ is the activation energy for the annihilation event and $\tau_0$ is the so-called pre-exponential factor.
%is related to the relative entropy of the initial and transition state as well as the flux through the transition state.
The two parameters, $\tau_0$ and $E_a$, can be estimated using the harmonic approximation to transition state theory 
(HTST) for magnetic degrees of freedom.\cite{Bessarab_2012,Bessarab_2013} 
It is based on an analysis of the multidimensional energy surface of the system, describing how the energy depends on the 
%where the dimensionality of the energy surface is equal to the number of 
angular variables 
%needed to 
specifying the direction of all the magnetic moments in the system.
In the case of interest, the antiskyrmion corresponds to a local minimum on the energy surface, 
whereas the homogeneous ferromagnetic phase corresponds to the global minimum. 
The activation energy for annihilation can be estimated as the highest rise in energy along the minimum energy path (MEP) 
connecting the antiskyrmion minimum to the ferromagnetic minimum.
The point of highest energy on the MEP corresponds to a first order saddle point on the energy surface.
The MEP can be found using the geodesic nudged elastic band method.\cite{Bessarab_2015}
The computational effort is reduced here by making use of prior 
knowledge about the shape of the MEP and by focusing only on a small part of the MEP near the maximum.
%(truncated MEP method)
\cite{Lobanov_2017}
 
% revised in resubmitted version 
In HTST, the pre-exponential factor, $\tau_0$, 
is related to the relative vibrational entropy of the initial and transition states as well as the flux through the transition state.
The vibrational entropy is estimated 
by approximating the energy surface in the vicinity of the local minimum and 
the saddle point as quadratic functions to obtain the frequencies of the vibrational modes.
The vibrational entropy of a state is then given by the product of the vibrational frequencies for that state.
Normally this involves evaluating the eigenvalues of the Hessian matrix\cite{Bessarab_2012}
but for large systems this becomes a challenging calculation.
Below, we briefly describe a more efficient method for evaluating $\tau_0$ that does not require the evaluation of 
the eigenvalues and makes it possible to carry out calculations for the large system studied here.
%  add in resubmitted manuscript
Additional information about the method can be found in Ref. \cite{Lobanov_2020}
The lowest couple of eigenvalues of the Hessian are still calculated explicitly
using the Lanczos method to ensure that only one negative eigenvalue is found at the obtained saddle point 
and to test for the possible presence of zero modes.

% ----------------------  Figure 2 ------------------------------
\begin{figure}[t!]
\begin{center}
\includegraphics[width=8 true cm]{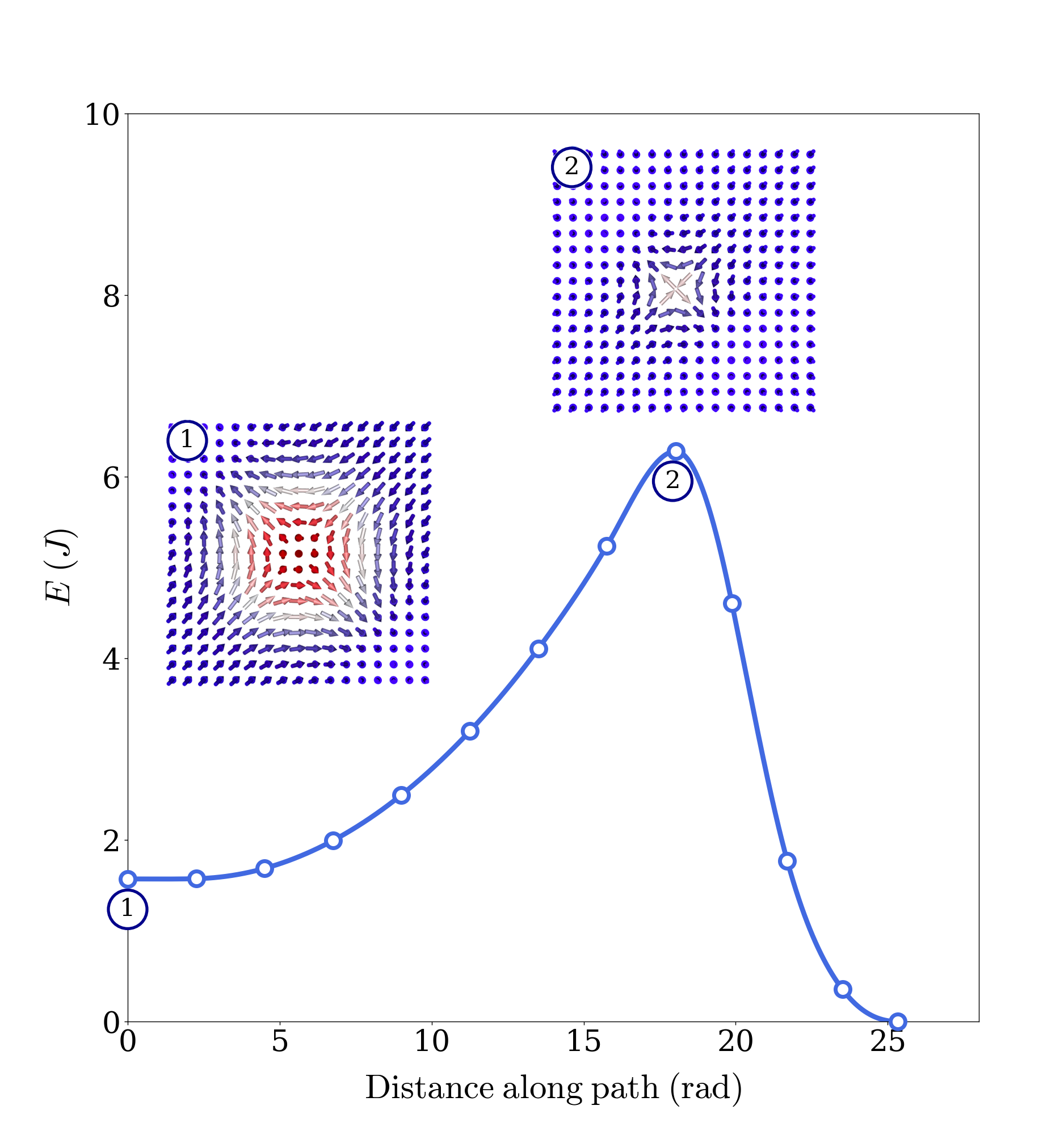}
\caption{\label{fig:2} 
Minimum energy path for the collapse of an antiskyrmion in a system when $N=1$, i.e. containing 45x45 spins.
The initial state and saddle point
spin configurations are shown in insets (only part of the simulated system is shown).
%  new in resubmitted manuscript
The final state at a distance of around 25 radians is the ferromagnetic state.
The distance along the path 
is the sum of the geodesic displacements corresponding to changes in the orientation of all the magnetic vectors in the system.
%gives the geodesic displacement summed over all the spins in the system.
}
\end{center}
\end{figure}
% ------------------------------------------------------------------

% ----------------------------------  Figure 3 ----------------------------------
\begin{figure}[b]
	\centering	
	\includegraphics[width=9 true cm]{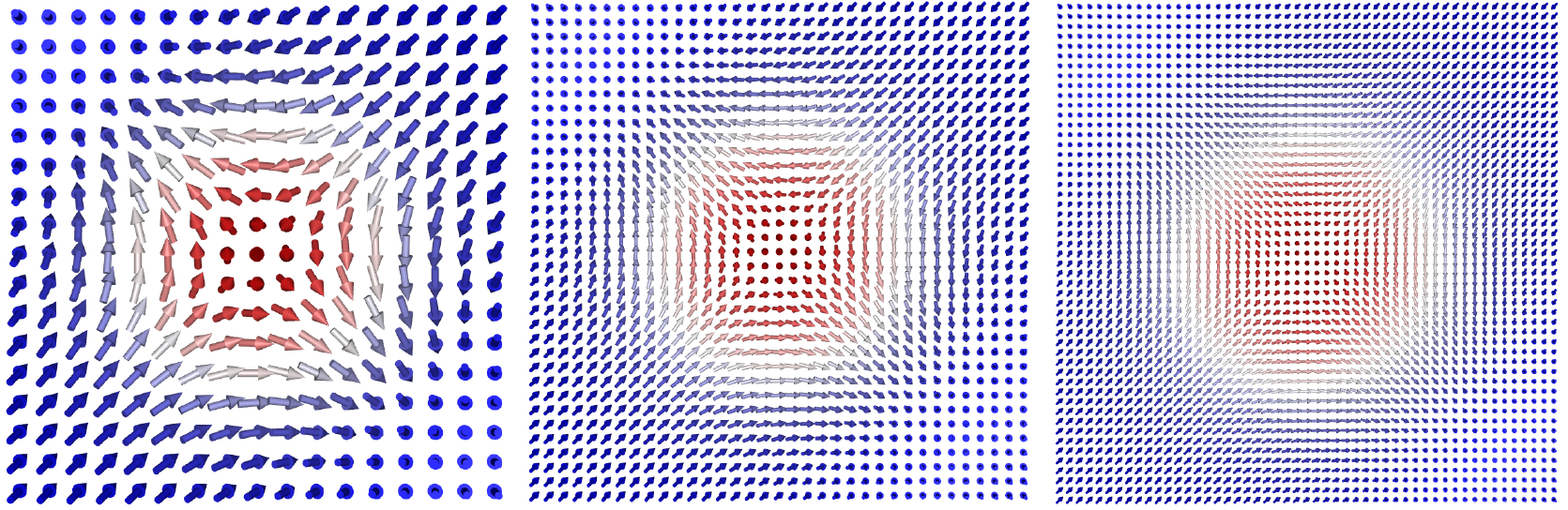}
	\caption{\label{fig:antiskyrmions} 
   Antiskyrmion configuration for $N=1$ (left),  $N=2$ (middle), and $N=3$ (right).
% Add in revised manuscript:
   The color (red vs. blue) indicates the direction of the out-of-plane component of the
magnetic vector and the intensity of the color indicates the magnitude.
    }
\end{figure}

% ----------------------------------------------------------------------------------

%An atomic scale simulation of such a large system is computationally challenging because of the large number of variables. In the calculations of the activation energy, the initial structures used in the relaxation to the antiskyrmion minimum and convergence to the saddle point configuration are obtained from results obtained for a smaller number of lattice sites. Calculations are carried out for systems with varying number of sites,  45$N$ x 45$N$, starting from  $N$=1, and eventually reaching the 900 x 900 lattice with spacing between spins corresponding to the lattice constant of the Mn–Pt–Sn inverse Heusler compound when N=20. In such a coarse graining procedure, the parameters in the Hamiltonian ($J$, $D$, $K$ and $\mu$), need to be rescaled in such a way that the antiskyrmion profile remains constant, i.e. $D_N=D_1 /N$, $K_N=K_1/N^2$, and $\mu_N= \mu_1/N^2$, whereas exchange parameter remains constant, $J_N=J_1$. Figure 3 shows the spin configurations corresponding to the first three values of $N$.

% revised version
An atomic scale simulation of such a large system is computationally challenging because of the large number of variables. 
In order to reduce computational effort we use the following scaling method. 
A sequence of calculations is carried out for systems with decreasing in-plane lattice constant, $a$, and increasing number of spins in such a way as to keep the values of the parameters corresponding to the micromagnetic model constant. Letting $a_N = a_1/N$ denote the in-plane lattice constant after $N$ iterations and referring to the relationship between micromagnetic model parameters, which are kept constant, and atomic lattice parameters, which depend on the lattice constant, the following scaling relationships are obtained:
$D_N=D_1/N$ and  $\mu_N=\mu_1/N^2$ while the exchange constant $J$ is unchanged as it is independent of $a$.
The number of spins is 45$N$ x 45$N$, starting from $N=1$, and eventually reaching the 900 x 900 when $N=20$ giving spacing between sites that corresponds to the lattice constant of the Mn--Pt--Sn inverse Heusler compound. The calculation for an MEP for a given $N$ starts from an initial guess obtained from the calculation with $N-1$. 
Fig. 3 shows the spin configurations corresponding to the first three values of $N$.

The pre-exponential factor is evaluated separately for each $N$ but without having to determine the eigenvalues of the Hessian as has
been done in previous calculations.\cite{Bessarab_2012,Bessarab_2013}
It is difficult to obtain high enough accuracy for the eigenvalues when the number of spins is so large.
Since the dipole-dipole interaction is not included here explicitly, only the nearest neighbors interact making it possible to
write the Hessian matrix in a block tri-diagonal form and thereby evaluate the determinant of the Hessian.
The pre-exponential factor is then evaluated from the determinant directly without having to determine the eigenvalues.  
It can be written as
\begin{equation}
\tau_0= \ \frac{2\pi}{\lambda \Omega_0}
\end{equation}
where the factor $ \Omega_0 $ is a ratio of determinants of the Hessian at the initial state minimum, $ H^{min} $, and at the saddle point, $ H^{sp} $,
i.e. 
\begin{equation}
\Omega_0 = {{ \sqrt{\det{H^{min}} } }\over {\sqrt{ \left| \det{H^{sp}} \right|}}}.
\end{equation}
It is connected with the ratio of the entropy of the initial state and the entropy of the transition state, both evaluated within the harmonic approximation. 
The factor $ \lambda $ is connected with the dynamics of the system through the transition state 
and is evaluated from a basis invariant expression as
\begin{equation}
\lambda=\sqrt{\frac{{\bf b}\cdot H^{sp} {\bf b}}{|\zeta|}},\quad {\rm with} \ \ 
{\bf b}=\frac{\gamma\zeta}{\mu} {\bf S}^{sp} \times {\bf e},
\end{equation}
where ${\bf S}^{sp}$ is the spin configuration at the saddle point,
$\zeta$ the negative eigenvalue of $H^{sp}$, 
$\bf e$ the corresponding eigenvector  
(a unit tangent vector to the MEP at the saddle point),
and 
$\gamma$ the gyromagnetic ratio.
This expression for $\lambda$ can be evaluated numerically even for large systems 
since it can be computed in spin-related basis without the evaluation of
the whole set of eigenvectors and eigenvalues of $H^{sp}$.\cite{Potkina_2020}
%Also, $\lambda$ can then be interpreted without referring to the dividing surface.
%Indeed, ${\bf b}\cdot H^{sp} {\bf b}$ is the value of the Hessian in the direction $\bf b$.
%The velocity is zero at ${\bf S}^{sp}$ since the gradient is zero, but the velocity at other points along the MEP is not zero,
%and in the vicinity of the saddle point the velocity
%is equal to $\bf b$ times the distance from the saddle point.
%This means that 
%the value of ${\bf b} \cdot H^{sp}{\bf b}$ can be interpreted as the curvature of the
%energy surface at the saddle point in the direction of $\bf b$.

\begin{figure}[t]
\begin{center}
\includegraphics[width=8 true cm]{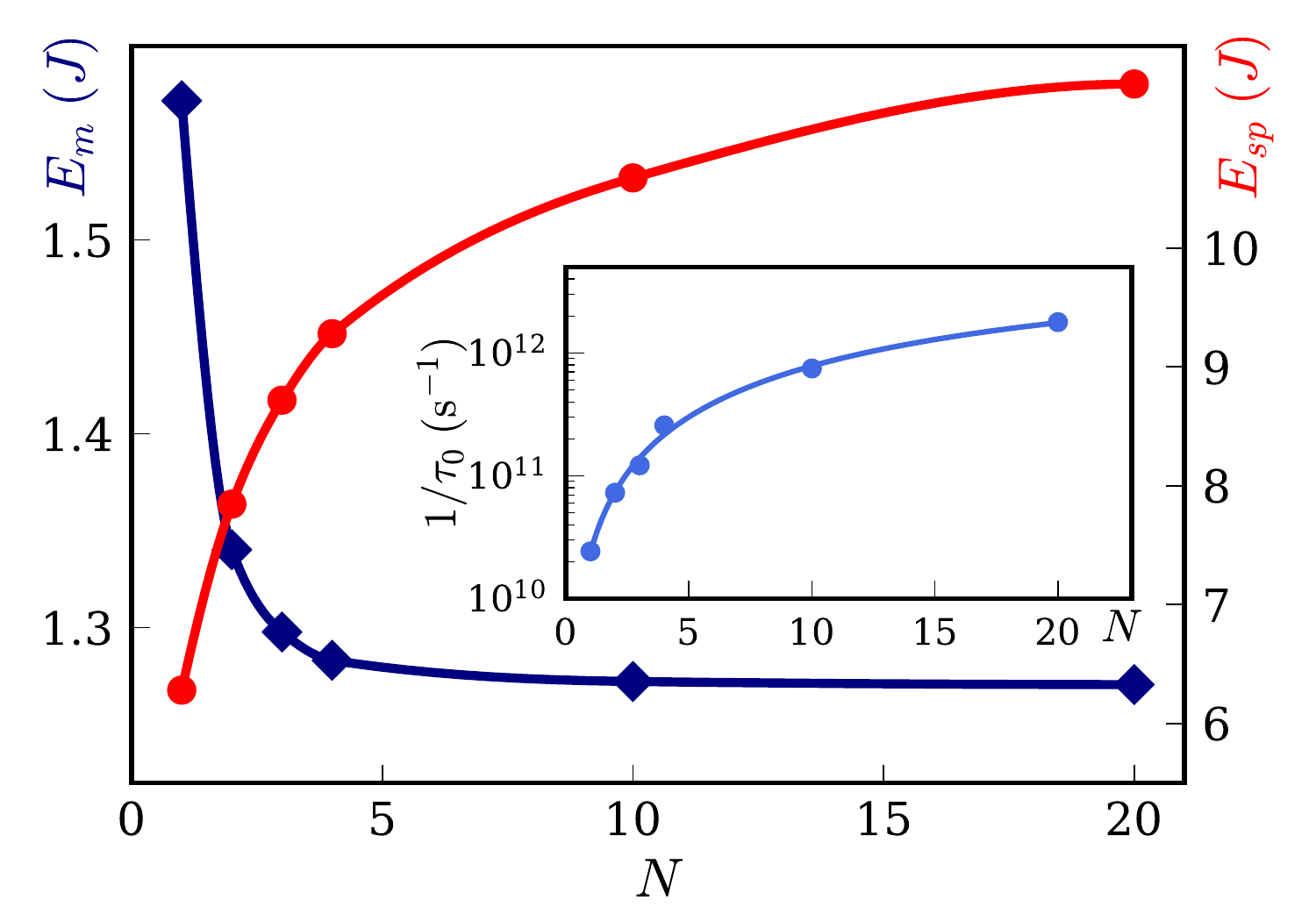}
\caption{\label{fig:4} 
Antiskyrmion energy at the local minimum, $E_{m}$ (blue line with diamonds), 
and at the saddle point for collapse, $E_{sp}$ (red line with circles), as a function of the scaling parameter $N$. 
The inset shows the pre-exponential factor in the Arrhenius rate law for antiskyrmion collapse as a function of $N$ assuming $J$ = 110 meV.   
The lattice parameter in the coarse grained simulation model corresponds that of the Mn--Pt--Sn inverse Heusler compound\cite{Nayak_2017} when $N=20$. 
}
\end{center}
\end{figure}

Figure~\ref{fig:4} shows the energy of the antiskyrmion with respect to the uniform ferromagnetic state 
and the energy of the saddle point as a function of the scaling parameter $N$.  
The energy at the local minimum corresponding to the antiskyrmion almost reaches a constant as the parameter $N$ is increased,
while the energy of the saddle point is still increasing at $N=20$.  
The lattice effects are still strong at $N=20$ and the Belavin-Polyakov limit\cite{Belavin_1975} 
of $4 \pi J$
has not yet been reached.
For the full 900 x 900 lattice corresponding to $N=20$, 
the antiskyrmion energy is found to be $E_{m}/J = 1.27$  and the saddle point energy $E_{sp}/J =11.38$. 
Thus, the energy barrier for the collapse of the antiskyrmion is $E_a/J = 10.11$. 
(See Ref. \cite{AnimationOfMEP} for an animation of the MEP for antiskyrmion collapse when $N=1$). 

% ----------------------  Figure 4 ------------------------------
% ---------------------- ---------------------- ----------------------

The largest uncertainty lies in the value of the exchange parameter, $J$. 
%In the micromagnetic modeling\cite{Nayak_2017}, the value was taken to be $J=1830$ meV.
The value of exchange stiffness used in the micromagnetic modeling of Nayak {\it et al.}\cite{Nayak_2017} corresponds to $J$=1830 meV
for the lattice representation.
Even for a much smaller value of $J$=110 meV
the activation energy for collapse is large,  {\it ca.}~1 eV.
%$J$=110 meV, is chosen rather than the one assumed in the micromagnetic modeling ($J=1830$)\cite{Nayak_2017}.
However, the pre-exponential factor turns out to have a typical value 
of 10$^{-12}$\,s (see inset in figure~\ref{fig:4}). 
Together, these values give a lifetime of a month at room temperature.
A larger value of $J$ would give even longer lifetime. 
This high stability is in good agreement with the reported experimental observations.\cite{Nayak_2017}
The reason for the long lifetime is the large activation energy for the collapse of the antiskyrmion in this material 
while the pre-exponential factor,  
which is related to entropic effects, 
%has a similar value as for 
has a value that is similar to what has been calculated previously for 
the collapse of small skyrmions.\cite{Bessarab_2018}

% ------------------------------------------------------------------------------

\section{Conclusions} 

Calculations of the lifetime of large but submicron scale antiskyrmions in Mn--Pt--Sn tetragonal Heusler material are presented. 
The results are consistent with the observed stability at room temperature in recent experiments\cite{Nayak_2017} 
and show that the long lifetime is due to large activation energy for collapse rather than entropic effects.
Since the atomic scale representation of this system requires roughly a million spins, the calculations are challenging
and are made possible by using a scaling approach to evaluate 
the activation energy and an improved method for evaluating the pre-exponential factor in the Arrhenius rate expression.
The latter is possible because the dipole-dipole interaction is not included here.  Previous studies using the micromagnetic approach
have shown that the dipole-dipole interaction makes antiskyrmions more stable with respect to skyrmions.\cite{Camosi_2018}
The rate of collapse cannot, however, be evaluated from micromagnetic 
simulations but we expect that the inclusion of dipole-dipole interaction would further increase the activation energy while not
affecting the value of the pre-exponential factor significantly.

% =========================================================
\ 

\vskip 1.5 true cm

\section*{Acknowledgments}

This work was supported by the Icelandic Research Fund, 
the Research Fund of the University of Iceland, 
and 
the Russian Science Foundation (Grant 19-42-06302).
%, the Cooperative Research Project Program at the Research Institute of Electrical Communication, 
%Tohoku University and by UNSW Science International Seed grant.
The calculations were carried out at the Icelandic Research High Performance Computing facility.

\vskip 1 true cm

%%%%%%%%%%%%%%%%%%%%%%%%%%%%%%%%%%%%%%%%%%%%%%%%%%%%%%%%%%%%%%%%%%%%%


\begin{thebibliography}{00}
\bibitem{AnimationOfMEP}
Animation of the MEP for collapse of an antiskyrmion is shown in AntiskyrmionCollapseN=1.gif.
\end{thebibliography}
\end{document}